\documentclass[a4paper]{mem}
\usepackage{natbib}
\usepackage{graphicx}
\idline{74}{9999}

\begin{document}
\title{$HST$ astrometry:\ the Galactic constant $\Theta_0/R_0$}
\author{L.\ R.\ Bedin\inst{1}, G.\ Piotto \inst{1}, 
        I.\ R.\ King\inst{2} \and J.\ Anderson\inst{2}}
\offprints{L.\ R.\ Bedin}
\mail{bedin@pd.astro.it}
\institute{Uni.\ di Padova, dip.\ di Astronomia, vic.\ 
           Osservatorio 2, I-35122 Padova, Italy
\and      Uni.\ of California at Berkeley, Berkeley, 
	  CA 94720-3411, USA.}
\abstract{From  multi-epoch   WFPC2/$HST$~ observations we present  
astrometric measurements of the   absolute motion of the  bulge stars.
The presence  of an extragalactic  point-source candidate allows us to
measure   the   difference   between   the    Oort  constants,   $A-B$
$=$$\Theta_0/R_0$.  We find:\  $\Theta_0/R_0$ $=$ 27.4 $\pm$ 1.8  $\rm
km~s^{-1}~$kpc$^{-1}$.
}
\authorrunning{L.\ R.\ Bedin et al.}
\titlerunning{The Galactic constant $\Theta_0/R_0$}
\maketitle
\section{Introduction}
In Bedin  et al.\ (2001), two $HST$  deep observations of the globular
cluster M4, separated by a time base line  of $\sim 5$ yrs, allowed us
to  obtain   a  pure  sample  of   main sequence stars    in M4.   The
identification of an extra-Galactic point source enables us to use the
proper motions of field stars (which were  junk in Bedin et al.\ 2001)
to measure a fundamental parameter of the Galaxy.

%
\section{Measure of the Constant $\Theta_0/R_0$}
M4 is a  globular cluster projected on  the edge of the Galactic bulge
($\ell\simeq-9^\circ$,   $b\simeq16^\circ$).  We  expect  only a small
number of  foreground disk stars in our  fields, but in the background
we look through the   edge of the bulge  at  a height of  $\sim2$ kpc.
Although at such  heights the density of the  bulge is rather low, the
volume we are probing  is sizable, so that we  see  a large number  of
bulge stars.
Their absolute proper motion (pm) is just the  reflection of the Sun's
angular  velocity  with respect to  that  point; from   that pm we can
derive  the value of the angular  velocity of the  LSR with respect to
the  Galactic   center, which  is  the  fundamental  Galactic-rotation
constant $A-B=\Theta_0/R_0$ (cf.\ Kerr \& Lynden-Bell 1986).

To  derive this value  we need to:\ (1)  find the mean distance of the
bulge stars whose motion we are observing, (2) correct the observed pm
for  the velocity of the Sun  with respect to the  LSR, and (3) relate
the corrected pm to the  angular velocity of  the LSR with respect  to
the Galactic center.

For the distance of  the bulge stars that we  are observing,  we assume
the following working hypotheses:
(1) Disk and halo stars are a negligible component of the field stars
      in our   M4 images,  i.e., the  field   stars  are mainly  bulge
      members.
(2) The bulge stars on our line of sight are part of a spherical
      spatial  distribution around the  Galactic center. 
(3) Our observations go deep enough that we do not loose stars on the
      far side of the bulge.
From these assumptions, it follows that we can express the distance of
the centroid of the bulge stars along our line of sight (we will refer
to it as the bulge) as a geometrical constant$\times$the distance of
the Sun from the Galactic center.
This distance is
$R=R_0\cos\ell\cos b.$
If we take $R_0=8.0$ kpc, then $R=7.6$ kpc.\\

Next, to link ($\Theta_0/R_0$) to the observables, and to estimate the
Solar corrections, we introduce the  following formulas. These express
the  pm observed in  the direction  $(\ell,~b)$  as a function  of the
velocity vector in a Galactic rest frame defined as
$(U_{{\rm abs}},V_{{\rm abs}},W_{{\rm abs}})$ $=$ $(U,V+\Theta_0,W)$ 
{\tiny 
\begin{equation}
\left \{
\begin{array}{l}
\mu_{\ell\cos{b}} = \frac{(U_{abs}^2+V_{abs}^2)^{1/2}\sin(\phi-\ell)}
{k R_0 \cos{\ell} \cos{b} },\\
\mu_{b} = \\
\frac{
( ((U_{abs}^2+V_{abs}^2)^{1/2}\cos(\phi-\ell))^2 + 
W_{abs}^2)^{1/2} \sin(\psi-b) }
{k R_0 \cos{\ell} \cos{b} },\\
\phi = \tan^{-1} \frac{ V_{abs} }{ U_{abs} },\\
\psi = \tan^{-1}\frac{W_{abs}}
{(U_{abs}^2+V_{abs}^2)^{1/2}\cos(\phi-\ell)},
\end{array}
\right .
\label{eq:mu2AB}
\end{equation}
}
where $k=4.74$ is the equivalent in  km/s of one astronomical unit per
tropical year.

At this point we  can correct our observations  of  the motion  of the
bulge for the  Sun's peculiar velocity,  and obtain the components due
exclusively to   the LSR circular  motion   around the Galactic center
(which is related to $\Theta_0/R_0$) 
{\tiny 
\begin{equation}
\left \{
\begin{array}{l}
\mu_{\ell\cos{b}}^{\rm LSR} =  
\mu_{\ell\cos{b}}^{\rm observed}-\mu_{\ell\cos{b}}^{\odot} \equiv X,~\\
\mu_b^{\rm LSR} = \mu_b^{\rm observed} - \mu_b^{\odot} \equiv Y,
\end{array}
\right .
\label{eq:mu3AB}
\end{equation}
}
where we introduced $X$ and $Y$ to be more concise.
In the case of the LSR we have
$(U_{abs},V_{abs},W_{abs})~=~(0,\Theta_0,0)$, and so from 
Eq.\ \ref{eq:mu2AB} we have
$\mu_{\ell\cos{b}}^{\rm LSR} = \frac{-1}{k\cos(b)} \Theta_0/R_0,$
$\mu_b^{{\rm LSR}} = \frac{\tan{b}\tan{\ell}}{k} \Theta_0/R_0, $
and combining them in quadrature, we get 
{\tiny 
\begin{equation}
\left \{
\begin{array}{l}
\Theta_0 / R_0 \pm \sigma_{\Theta_0/R_0} = 
F \sqrt{X^2+Y^2} \pm F\sigma ~\\
F = k \cos{b}+[1+\sin^2{b}\tan^2{\ell}]^{-1/2} \\
\sigma=\sqrt{\sigma_X^2+\sigma_Y^2} \\
\end{array}
\right .
\label{eq:mu4AB}
\end{equation}
}
%
\begin{figure}
\centering
\includegraphics[height=6.0cm]{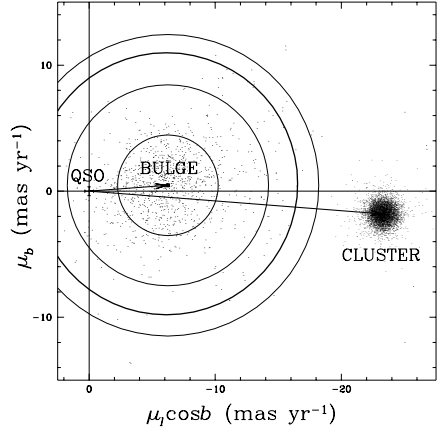}
\caption{Vector-point diagram of all the independent  measurements  of 
the  in Galactic  pms.  The  arrows indicates the  mean motion  of the
cluster and the bulge with respect to an extragalactic source.  }
\label{fig:bulgeAB} 
\end{figure}
%
Figure \ref{fig:bulgeAB} shows the  pms in Galactic coordinates.   The
origin has  been set at the  extra-galactic  point source labeled QSO.
We drew a heavy circle at  a radius of  4 times the semi-interquartile
distance of the field stars from their median position (for stars with
pm larger than 5 mas/yr with respect to the cluster mean).  We assumed
the stars  inside the circle  to belong to  the same distribution, and
calculated the mean from them.
The $\sigma$ has  been taken to  be the 68.27$^{th}$ percentile of the
distribution of sizes of the pms with respect to the mean.
In Fig.\   \ref{fig:bulgeAB}   the  three thin   circles  show   1, 2,
3$\sigma$.
{\it This is the mean absolute motion of the bulge} and it is shown as an
arrow in Fig.\  \ref{fig:bulgeAB}. 
With Eq.\ \ref{eq:mu2AB}   we    can estimate  the  Solar   correction
adopting:\ a Solar motion, and a
Galactocentric   distance; 
consequently from Eq.\ \ref{eq:mu3AB} we can estimate X and Y.  

The  absolute motion of  the  bulge---corrected for the Sun's peculiar
motion---allows  us to   get a direct   estimate of  the Oort-constant
difference $(A-B)$, which is related to the transverse velocity of the
LSR ($\Theta_0$) and its Galactocentric distance ($R_0$), according to
Eq.\ \ref{eq:mu4AB}, by  $(A-B) \pm \sigma_{(A-B)}$ $=$
$\Theta_0/R_0 \pm \sigma_{\Theta_0/R_0}$ $=$ 
$27.6 \pm 1.7$ ${\rm km/s~{\rm kpc.}}$
The  quoted error  is internal  and corresponds  to an uncertainty  of
7\%.
\bibliographystyle{aa}

\end{document}